\newcommand{\roughly}[1]%
    {{\mathrel{\raise.3ex\hbox{$#1$\kern-.75em\lower1ex\hbox{$\sim$}}}}}

\documentclass[12pt]{article}
\usepackage{paper2e}
\usepackage{latexsym}
\usepackage{graphicx}

\begin{document}
\begin{titlepage}

\preprint{IPMU09-0081}

\title{Caustic avoidance in Ho\v{r}ava-Lifshitz gravity}

\author{Shinji Mukohyama}

\address{
Institute for the Physics and Mathematics of the Universe (IPMU)\\ 
The University of Tokyo\\
5-1-5 Kashiwanoha, Kashiwa, Chiba 277-8582, Japan
}

\begin{abstract}

 There are at least four versions of Ho\v{r}ava-Lishitz gravity in the
 literature. We consider the version without the detailed balance
 condition with the projectability condition and address one aspect of
 the theory: avoidance of caustics for constant time hypersurfaces. We
 show that there is no caustic with plane symmetry in the absence of
 matter source if $\lambda\ne 1$. If $\lambda=1$ is a stable IR fixed
 point of the renormalization group flow then $\lambda$ is expected to
 deviate from $1$ near would-be caustics, where the extrinsic curvature 
 increases and high-energy corrections become important. Therefore, the
 absence of caustics with $\lambda\ne 1$ implies that caustics cannot
 form with this symmetry in the absence of matter source. We argue that 
 inclusion of matter source will not change the conclusion. We also
 argue that caustics with codimension higher than one will not form
 because of repulsive gravity generated by nonlinear higher curvature
 terms. These arguments support our conjecture that there is no caustic
 for constant time hypersurfaces. Finally, we discuss implications to
 the recently proposed scenario of ``dark matter as integration
 constant''. 
\end{abstract}

\end{titlepage}

\section{Introduction}
\label{sec:introduction}

Ho\v{r}ava recently proposed a power-counting renormalizable theory of
gravitation~\cite{Horava:2009uw,Horava:2008ih}. Since one of the most
important aspects of the theory is a Lifshitz-type anisotropic scaling,
it is often called Ho\v{r}ava-Lifshitz gravity. Various aspects of this
theory were investigated~\cite{Takahashi:2009wc}-\cite{Park:2009zr}.

Actually, in the literature there are at least four versions of the
theory: with/without the detailed balance condition; and with/without
the projectability condition. Among them, only the version without the  
detailed balance condition with the projectability condition has a
potential to be theoretically consistent and cosmologically viable
while there still are many unsolved issues. Ho\v{r}ava's original
proposal~\cite{Horava:2009uw} was with the projectability condition and
with/without the detailed balance condition~\footnote{Ho\v{r}ava put
much emphasis on the detailed balance condition but considered it as
just a way to reduce the number of independent coupling constants.}.

The detailed balance condition restricts the form of potential in the
$4$-dimensional Lorentzian action to a specific form in terms of a
$3$-dimensional Euclidean theory. From cosmological viewpoint, this
condition leads to obstacles~\cite{Nastase:2009nk,Calcagni:2009qw} and
thus must be abandoned.

On the other hand, the projectability condition stems from the
fundamental symmetry of the theory, i.e. the foliation-preserving 
diffeomorphism invariance, and thus must be respected. The
foliation-preserving diffeomorphism consists of the $3$-dimensional
spatial diffeomorphism and the space-independent time
reparametrization. Since the lapse function is essentially the gauge
degree of freedom associated with the time reparametrization, it is
natural to restrict it to be space-independent. This is the
projectability condition, and one can easily show that the
foliation-preserving diffeomorphism preserves this condition. The
projectability condition implies that the Hamiltonian constraint is not
a local equation satisfied at each spatial point but an equation
integrated over a whole space. Abandoning the projectability condition
and imposing a local version of the Hamiltonian constraint would result
in phenomenological obstacles~\cite{Charmousis:2009tc} and theoretical
inconsistencies~\cite{Li:2009bg}. Those problems disappear once the
projectability condition is respected and if only the global Hamiltonian
constraint is imposed (see, for example, section 5 of
\cite{Mukohyama:2009mz}). Note that Ho\v{r}ava's original proposal was
with the projectability condition.

For these reasons, in the present paper we restrict our attention to the
version without the detailed balance condition with the projectability
condition. We suppose that the dynamical critical exponent $z$ in the
ultraviolet (UV) is equal to or larger than $3$ since power-counting
(super-)renormalizability requires it. To the best of the author's
knowledge, no explicit inconsistency has been found against this
version~\footnote{Ref.~\cite{Blas:2009yd} argued that the version with
the projectability condition is also inconsistent, based on the
following two claims: (i) constant-time hypersurfaces form caustics;
(ii) the scalar graviton gets strongly coupled at all scales in
Minkowski background even away from $\lambda=1$. Actually, these claims
are not correct. See the last paragraph of subsection
\ref{subsec:highcodim}, and the fourth-to-the-last and third-to-the-last
paragraphs of section~\ref{sec:darkmatter} of the present paper.}.

This version of Ho\v{r}ava-Lifshitz gravity, with general $z$, has an
interesting cosmological consequence even in the infrared (IR). The
global Hamiltonian constraint is less restrictive than its local
version, and allows a richer set of solutions than in general
relativity. Actually, a component which behaves like pressureless dust 
emerges as an ``integration constant'' of dynamical equations and
momentum constraint equations~\cite{Mukohyama:2009mz}. As a result,
classical solutions to the infrared limit of Ho\v{r}ava-Lifshitz gravity
can mimic solutions to general relativity plus cold dark matter. We
shall discuss more about the ``dark matter'' in
Sec.~\ref{sec:darkmatter}.

Also in the UV, cosmology based on Ho\v{r}ava-Lifshitz gravity has
number of interesting properties. The anisotropic scaling with the
dynamical critical exponent $z=3$ leads to a new mechanism of generating
scale-invariant cosmological perturbations~\cite{Mukohyama:2009gg}. The
nonlinear higher spatial curvature terms lead to regular bounce
solutions in the early
universe~\cite{Calcagni:2009ar,Brandenberger:2009yt}. It was also
suggested that nonlinear higher spatial curvature terms might make the 
flatness problem milder than in general
relativity~\cite{Kiritsis:2009sh}. Since these mechanisms do not rely on 
the detailed balance condition or a local version of the Hamiltonian
constraint, they can be applied to the version we are considering now.

Now, while cosmological consequences are interesting and it is
worthwhile exploring further, one has to be aware that there are many
fundamental issues to be addressed in the future. First of all,
renormalizability beyond power-counting has not been proven. Second, the
renormalization group (RG) flow of various coupling constants has not
been investigated. In particular, recovery of general relativity in the
IR relies on the assumption (or hope) that the parameter $\lambda$ (see
the next section) flows to $1$ in the IR. Without knowing the condition
for this behavior to be realized, we cannot be sure about recovery of
general relativity in the IR. Third, this theory has not yet been
intended to be a part of unified theory. Clearly, further developments
or/and embedding into a ``bigger'' theory (or other way around) is
needed. In particular, since the ``limit of speed'' is subject to the RG
flow~\cite{Iengo:2009ix}, we need a new idea to ensure that different
species including those in the standard model of particle physics are
somehow related to each other so that their ``limits of speed'' agree
with the ``velocity of light'' within experimental limits~\footnote{See
e.g. refs.~\cite{Coleman:1998ti,Jacobson:2001tu,Moore:2001bv} for tight
experimental limits on Lorentz violation.}. Fourth, an analogue of the
Vainshtein effect~\cite{Vainshtein:1972sx} must be investigated for the
scalar graviton in the limit $\lambda\to 1$. Unlike general relativity,
Ho\v{r}ava-Lifshitz gravity has not only a tensor graviton with two
polarizations but also a scalar graviton. The dynamics of the scalar
graviton and its fate in the IR must be investigated in details. Since
the time kinetic term (together with gradient terms) of the scalar
graviton vanishes in the $\lambda\to 1$
limit~\cite{Horava:2009uw,Horava:2008ih}, one has to take into account
nonlinear interactions to see if it really decouples.

Aside from those issues, there is another important question. In general
relativity, flow of pressureless dust generically forms caustics. Thus,
one might expect that the flow of ``dark matter as integration
constant'' would also form caustics. If this were the case then constant 
time hypersurfaces would develop singularities since the flow of ``dark 
matter'' is orthogonal to constant time hypersurfaces. This would be a
disaster for Ho\v{r}ava-Lifshitz gravity with the projectability
condition since the constant time hypersurfaces have physical meaning in
this theory. Indeed, if caustics formed then the extrinsic curvature
would diverge.

In this paper, we shall argue that this naive expectation is not
correct and conjecture that there is no caustic for constant time
hypersurfaces. While the proof of this conjecture is beyond the scope of
this paper, we shall provide supporting arguments. In particular, we
shall show that there is no caustic with plane symmetry in the absence
of matter source if $\lambda\ne 1$. Since near would-be caustics 
the system enters the UV regime and $\lambda$ is expected to deviate
from $1$ via RG flow~\footnote{Here, it is assumed that the theory is
renormalizable. It is also assumed that $\lambda=1$ is a stable IR fixed
point of RG flow. In this case, by reversing the direction of the RG
flow and going towards the UV, $\lambda$ should deviate from
$1$. \label{footnote:deviationoflambda}}, this implies that caustics
with codimension one do not form in the absence of matter source. We
shall argue that inclusion of matter source will not change the
conclusion. Since caustics with lower codimensions are more
difficult to bounce than those with higher
codimensions~\cite{ArkaniHamed:2005gu}, this result provides a strong
support for the conjecture. We shall also argue that caustics with
codimension higher than one will not form because of repulsive gravity
generated by nonlinear higher curvature terms. Finally, we shall discuss
implications to the recently proposed scenario of ``dark matter as
integration constant''.

The rest of this paper is organized as follows. In
Sec.~\ref{sec:basic-equations} we summarize basic equations in
Ho\v{r}ava-Lifshitz gravity with the projectability condition. In
Sec.~\ref{sec:no-caustic} we shall argue that caustics do not
form. Sec.~\ref{sec:darkmatter} is devoted to discussion of implications
to the ``dark matter as integration constant'' scenario.

\section{Basic equations}
\label{sec:basic-equations}

The basic quantities in the theory are the spatial metric $g_{ij}$, the
shift vector $N^i$ and the lapse function $N$. While $g_{ij}$ and $N^i$
can depend on both space and time coordinates, $N$ can depend only on
$t$. The fundamental symmetry of the theory is the invariance under the
foliation-preserving diffeomorphism:
%
\begin{equation}
 t \to \tilde{t}(t), \quad x^i \to \tilde{x}^i(t,x).
\end{equation}
Under the infinitesimal transformation
%
\begin{equation}
 \delta t = f(t), \quad \delta x^i = \zeta^i(t,x),
\end{equation}
$g_{ij}$, $N^i$ and $N$ transform as
%
\begin{eqnarray}
 \delta g_{ij} & = & f\partial_t g_{ij} + {\cal L}_{\zeta}g_{ij}
  \nonumber\\
 \delta N^i & = & \partial_t (N^i f) + \partial_t \zeta^i
  + {\cal L}_{\zeta}N^i, \nonumber\\
 \delta (N_i) & = & \partial_t (N_i f) + g_{ij}\partial_t \zeta^j
  + {\cal L}_{\zeta}N_i, \nonumber\\
 \delta N & = & \partial_t(N f). 
  \label{eqn:infinitesimal-tr}
\end{eqnarray} 
Thus, $N$ remains independent of spatial coordinates after
transformation. In the IR, where $dt$ and $dx^i$ have the same scaling
dimension, it makes sense to assemble $g_{ij}$, $N^i$ and $N$ into a
$4$-dimensional metric in the ADM form: 
%
\begin{equation}
 ds^2 = -N^2dt^2 + g_{ij}(dx^i+N^idt)(dx^j+N^jdt). 
\end{equation}

The action is 
%
\begin{equation}
 I = I_g+I_m, \quad 
  I_g = \frac{M_{Pl}^2}{2}
  \int dt dx^3N\sqrt{g} (K^{ij}K_{ij}-\lambda K^2+R+L_{z>1}),
\end{equation}
where
%
\begin{equation}
 K_{ij} = \frac{1}{2N}(\partial_t g_{ij}-D_iN_j-D_jN_i), \quad
  K = g^{ij}K_{ij}, 
\end{equation}
$D_i$ is the covariant derivative compatible with $g_{ij}$, $R$ is the
Ricci scalar of $g_{ij}$, $L_{z>1}$ represents higher spatial curvature
terms and $I_m$ is the matter action. Here, we have rescaled the time
coordinate so that the coefficients of $K^{ij}K_{ij}$ and $R$ agree. The
cosmological constant term will be included in matter action if
necessary. Note that not only the gravitational action $I_g$ but also
the matter action $I_m$ should be invariant under the
foliation-preserving diffeomorphism.

By variation of the action with respect to $N(t)$, we obtain the
Hamiltonian constraint 
%
\begin{equation}
 H_{g\perp}+H_{m\perp}=0, \label{eqn:HamiltonianConstraint}
\end{equation}
%
\begin{eqnarray}
 H_{g\perp} & \equiv & -\frac{\delta I_g}{\delta N}
  = \int dx^3\sqrt{g} {\cal H}_{g\perp}, 
  \quad
  {\cal H}_{g\perp} = \frac{M_{Pl}^2}{2}
  (K^{ij}p_{ij}-R-L_{z>1}),
  \nonumber\\
 H_{m\perp} & \equiv & -\frac{\delta I_m}{\delta N}
  = \int dx^3\sqrt{g}\ T^{\perp}_{\perp}, 
  \quad T^{\perp}_{\perp} = T_{\mu\nu}n^{\mu}n^{\nu}.
\end{eqnarray}
Here, $p_{ij}$ and $n^{\mu}$ are defined as 
%
\begin{equation}
 p_{ij} \equiv K_{ij} - \lambda Kg_{ij},
\end{equation}
and
%
\begin{equation}
 n_{\mu}dx^{\mu} = -Ndt, \quad
  n^{\mu}\partial_{\mu}= \frac{1}{N}(\partial_t-N^i\partial_i). 
  \label{eqn:unitnormal}
\end{equation}
Variation with respect to $N^i(t,x)$ leads to the momentum constraint 

%
\begin{equation}
 {\cal H}_{g i}+{\cal H}_{m i}=0, \label{eqn:MomentumConstraint}
\end{equation}
%
\begin{eqnarray}
 {\cal H}_{g i} & \equiv & -\frac{1}{\sqrt{g}}\frac{\delta I_g}{\delta N^i}
  = -M_{Pl}^2D^jp_{ij}, \nonumber\\
 {\cal H}_{m i} & \equiv & -\frac{1}{\sqrt{g}}\frac{\delta I_m}{\delta N^i}
  = T_{i\mu}n^{\mu}. 
\end{eqnarray}

As in general relativity, the gravitational action can be written as the
sum of kinetic terms and constraints up to boundary terms:  
%
\begin{equation}
 I_g = \int dt dx^3
  \left[\pi^{ij}\partial_tg_{ij}-N^i{\cal H}_{gi}\right]
  - \int dt NH_{g\perp} + (\mbox{boundary terms}),
\end{equation}
where $\pi^{ij}$ is momentum conjugate to $g_{ij}$ given by 
%
\begin{equation}
 \pi^{ij} \equiv \frac{\delta I_g}{\delta (\partial_tg_{ij})}
  = M_{Pl}^2\sqrt{g}p^{ij}, \quad
  p^{ij} \equiv g^{ik}g^{jl}p_{kl}. 
\end{equation}
The Hamiltonian corresponding to the time $t$ is the sum of constraints
and boundary terms as
%
\begin{equation}
 H_g[\partial_t] = NH_{g\perp} + \int dx^3 N^i{\cal H}_{gi}
  + (\mbox{boundary terms}). 
\end{equation}

Finally, by variation with respect to $g_{ij}(t,x)$, we obtain dynamical
equation 
%
\begin{equation}
 {\cal E}_{g ij}+{\cal E}_{m ij}=0, \label{eqn:DynamicalEquation}
\end{equation}
%
\begin{eqnarray}
 {\cal E}_{g ij} & \equiv & g_{ik}g_{jl}\frac{2}{N\sqrt{g}}
  \frac{\delta I_g}{\delta g_{kl}}, \nonumber\\
 {\cal E}_{m ij} & \equiv & g_{ik}g_{jl}\frac{2}{N\sqrt{g}}
  \frac{\delta I_m}{\delta g_{kl}}
  = T_{ij}. 
\end{eqnarray} 
The explicit expression for ${\cal E}_{g ij}$ is given by 
%
\begin{eqnarray}
 {\cal E}_{g ij} & = & M_{Pl}^2
  \left[
  -\frac{1}{N}(\partial_t-N^kD_k)p_{ij} - Kp_{ij} + 2K_i^kp_{kj}
  \right.\nonumber\\
 & & \left.
  + \frac{1}{N}(p_{ik}D_jN^k+p_{jk}D_iN^k)
  + \frac{1}{2}g_{ij}K^{kl}p_{kl} - G_{ij}
	\right]
 + {\cal E}_{z>1 ij},
\end{eqnarray} 
where ${\cal E}_{z>1 ij}$ is the contribution from $L_{z>1}$ and
$G_{ij}$ is Einstein tensor of $g_{ij}$.

The invariance of $I_{\alpha}$ under the infinitesimal transformation 
(\ref{eqn:infinitesimal-tr}) leads to the following conservation
equations, where $\alpha$ represents $g$ or $m$. 
%
\begin{eqnarray}
 N\partial_t H_{\alpha\perp}
  + \int dx^3\left[ N^i\partial_t(\sqrt{g}{\cal H}_{\alpha i})
  +\frac{1}{2}N\sqrt{g}{\cal E}_{\alpha}^{ij}\partial_tg_{ij}\right]
  & = & 0, \nonumber\\
 \frac{1}{N}(\partial_t-N^jD_j){\cal H}_{\alpha i}
  + K{\cal H}_{\alpha i}
  - \frac{1}{N}{\cal H}_{\alpha j}D_iN^j
  - D^j{\cal E}_{\alpha ij} & = & 0.
  \label{eqn:conservation}
\end{eqnarray} 

\section{Absence of caustics}
\label{sec:no-caustic}

The vector $n^{\mu}$ defined in (\ref{eqn:unitnormal}) has unit norm and
is orthogonal to constant time hypersurfaces. Since the lapse function
$N$ depends only on time, $n^{\mu}$ follows the geodesic equation. 
%
\begin{equation}
 n^{\mu}\nabla_{\mu}n_{\nu} =
  n^{\mu}\nabla_{\nu}n_{\mu} =
  \frac{1}{2}\partial(n^{\mu}n_{\mu}) = 0. 
  \label{eqn:geodesic}
\end{equation}

In general relativity, a congruence of geodesics would generically form
caustics. As an example, let us consider a congruence of geodesics
orthogonal to the hypersurface $t=T(x)$ in Minkowski spacetime
$ds^2=-dt^2+dx^2+dy^2+dz^2$. By introducing a new time coordinate $\tau$
as the proper time along each geodesic and a new spatial coordinate $X$
as the value of $x$ at the intersection of each geodesic with the
hypersurface $t=T(x)$, the Minkowski metric is rewritten as 
%
\begin{equation}
 ds^2 = -d\tau^2 + a^2(\tau,X)dX^2 + dy^2 + dz^2,
  \quad 
  a(\tau,X) = \sqrt{1-[T'(X)]^2}\left[1-\frac{\tau}{\tau_c(X)}\right], 
\end{equation}
where 
%
\begin{equation}
 \tau_c(X) = \frac{[1-T'(X)^2]^{3/2}}{-T''(X)}. 
\end{equation}
The metric component $a^2(\tau,X)$ vanishes at finite proper time
$\tau=\tau_c(X)$ and, thus, the congruence of geodesics form
caustics. In general relativity with a certain energy condition, it is
easy to show that a congruence of geodesics forms caustics in more
general situations essentially because gravity is attractive.

On the other hand, for the vector $n^{\mu}$ in the Ho\v{r}ava-Lifshitz
gravity, higher curvature terms become important near (would-be)
caustics and provide negative effective energy and repulsive
gravity. Also, $\lambda$ should deviate from $1$ by RG flow (see
footnote \ref{footnote:deviationoflambda}) and, thus, the kinetic terms
also contribute differently from general relativity.

\subsection{Codimension higher than one}
\label{subsec:highcodim}

For codimension higher than one, the spatial curvature of the constant
time hypersurface increases near (would-be) caustics. The system enters
the non-relativistic regime and higher spatial curvature terms become
important. Among them, highest order terms (e.g. curvature cubic terms
in the case of $z=3$) generate the strongest restoring force. As in some
early universe
models~\cite{Calcagni:2009ar,Kiritsis:2009sh,Brandenberger:2009yt}, we
expect that the (would-be) caustics should bounce at short distance
scales if codimension is higher than one. Note that this is not because 
of deviation from geodesics~\footnote{
In the case of ghost condensate~\cite{ArkaniHamed:2003uy}, the
derivative of the scalar field responsible for the condensate deviates
from geodesics because of higher derivative
terms~\cite{ArkaniHamed:2005gu}. On the other hand, in
Ho\v{r}ava-Lifshitz gravity the vector $n^{\mu}$ always satisfies the
geodesic equation (\ref{eqn:geodesic}).} but because of repulsive
gravity at short distances. In general relativity, congruence of
geodesics would form caustics because gravity is attractive. On the
other hand, for the vector $n^{\mu}$ in Ho\v{r}ava-Lifshitz gravity,
higher curvature terms become important near the (would-be) caustics and
provide repulsive gravity and thus bounce.

For an odd $z$, the sign of the highest nonlinear spatial curvature term
can in principle change. Nonetheless, as far as a contracting region has
a finite volume, the leading (would-be) divergence in the spatial
curvature at late time is expected to be positive. For this reason, even
with an odd $z$, we expect would-be caustics to bounce eventually. On
the other hand, numerical confirmation of this kind of behavior probably
requires a rather wide dynamic range since we have to wait until the
finiteness of the contracting region becomes important. Thus, for
numerical purposes it is probably easier to consider a large enough,
even $z$. We hope to perform numerical analysis of the bouncing behavior
in the future.

One might worry about the fact that the anisotropic scaling in the UV
leads to the scaling $\propto 1/a^{z+3}$ for radiation energy 
density~\cite{Mukohyama:2009zs}. For $z=3$, this might cause
difficulties for bouncing cosmology with FRW
(Friedmann-Robertson-Walker) spacetime since the radiation energy
density scales in the same way as the $z=3$ higher curvature terms and
has the opposite sign. On the other hand, in the present situation, the
contracting region has just a finite volume and thus radiation does not
have to be comoving with the vector $n^{\mu}$. Indeed, while the ``dark
matter'' is pressure-less and the vector $n^{\mu}$ follows geodesics,
the large pressure ($P_{rad}=(z/3)\rho_{rad}$ in the UV) acts as extra
repulsive force for radiation and prevents radiation from following
geodesics during inhomogeneous contraction. Thus, radiation can easily
diffuse and grows much more slowly than the $z=3$ curvature terms. For
this reason, for $z=3$, radiation does not cause difficulties. For
$z>3$, highest nonlinear curvature terms ($\propto 1/a^{2z}$) grows
faster than radiation energy density even in FRW spacetime and, thus,
radiation does not prevent the highest curvature terms from acting as
restoring forces.

Therefore, unless the whole universe contracts, (would-be) caustics with
codimension higher than one should bounce due to higher curvature 
terms. If the universe is completely homogeneous and flat, i.e. if the
universe is a flat FRW spacetime, then higher spatial curvature terms
vanish classically. Hence, a contracting flat FRW does not bounce
classically. However, quantum mechanically, there must be fluctuations
and nonlinear higher spatial curvature terms must acquire non-vanishing 
expectation values. Those fluctuations grow as the universe
contracts. Therefore, if $z$ is large enough and perhaps if $z$ is even
(so that the highest nonlinear spatial curvature terms always act as a
restoring force) then a contracting flat FRW universe might also bounce
after all. Further investigation of this issue is worthwhile.

Note that the bounce is provided by nonlinear terms. Therefore, if we
analyzed behaviors of the vector $n^{\mu}$ and constant time
hypersurfaces without including those nonlinear terms then we would not
be able to see the bounce and would instead see caustics forming. It is
likely that this is closely related to instabilities of linear
perturbation found in \cite{Sotiriou:2009bx}. It is also important to
include backreactions of the higher spatial curvature terms to the
geometry since, as already stated, the bounce is not due to deviation
from geodesics but due to repulsive gravity at short distances. If we
did not take into account backreactions to the geometry, one would
simply conclude formation of caustics~\cite{Blas:2009yd}. Without taking
into account nonlinear terms and backreactions to the geometry, we would
never be able to describe the system properly.

\subsection{Codimension one}

On the other hand, for codimension one, higher spatial curvature terms
do not help. This is in accord with the observation in
\cite{ArkaniHamed:2005gu} that caustics with lower codimensions are more
difficult to bounce than those with higher codimensions. In order to see
it, let us consider the following ansatz with plane symmetry. 
%
\begin{equation}
 N=\alpha(t), \quad N^i\partial_i = \beta(t,x)\partial_x, \quad 
  g_{ij}dx^idx^j = a(t,x)^2dx^2 + dy^2 + dz^2.
\end{equation}
We have the foliation-preserving diffeomorphism 
%
\begin{equation}
 t\to \tilde{t}(t), \quad x\to \tilde{x}(t,x), \quad 
  y\to y, \quad z\to z. 
\end{equation}
By using this symmetry, we can set $\alpha=1$ and $a=1$. The ansatz is
now reduced to 
%
\begin{equation}
 N=1, \quad N^i\partial_i = \beta(t,x)\partial_x, \quad 
  g_{ij}dx^idx^j = dx^2 + dy^2 + dz^2.
  \label{eqn:ansatz}
\end{equation}
We still have the residual symmetry
%
\begin{equation}
 t \to t+t_0, \quad x\to x+ x_0(t),
  \label{eqn:residual-symmetry}
\end{equation}
where $t_0$ is a constant and $x_0$ is a function of $t$. It is now
evident that higher spatial curvature terms do not help since the
spatial metric is flat.

Actually, what prevents caustics with codimension one from forming is
deviation of $\lambda$ from $1$. In order to recover general relativity
in the IR, $\lambda=1$ must be an IR fixed point of RG flow. However,
near (would-be) caustics, the system enters the UV regime and $\lambda$
should deviate from $1$ by RG flow (see footnote
\ref{footnote:deviationoflambda}). In the following, we shall show that
there is no caustics with codimension one if $\lambda\ne 1$.

Ideally speaking, it is appropriate to take into account fully quantum
mechanical effects to analyze behavior of the system near would-be
caustics. However, at this stage where renormalizability (beyond
power-counting) and RG flow of the theory are not yet understood, it is
not easy to perform fully quantum mechanical analysis. For this reason,
in this paper, we partially take into account quantum effects by
including the scale-dependence of coupling constants (especially,
deviation of $\lambda$ from $1$) in classical equations of motion.

As already argued in the previous subsection, radiation energy density
does not become significant near (would-be) caustics simply because
radiation does not have to be comoving with the vector $n^{\mu}$ and
thus can diffuse. The same applies to other forms of matter
fields. For this reason, in the following we ignore matter fields and
set $T_{\mu\nu}=0$.

For the ansatz (\ref{eqn:ansatz}), it is easy to show that
%
\begin{eqnarray}
 {\cal H}_{g x} & = & -(\lambda-1)M_{Pl}^2\beta'', 
  \nonumber\\
 {\cal E}_{g xx} & = & (\lambda-1)M_{Pl}^2
  \left[-\dot{\beta}'+\beta\beta''+\frac{1}{2}(\beta')^2\right], 
  \nonumber\\
 {\cal E}_{g yy} & = &  {\cal E}_{g zz}
  = \lambda M_{Pl}^2
  \left[-\dot{\beta}'+\beta\beta''+\frac{1}{2}(\beta')^2\right]
  + \frac{M_{Pl}^2}{2}(\beta')^2,
\end{eqnarray}
where an overdot and a prime denote derivatives with respect to $t$ and
$x$, respectively. With $\lambda\ne 1$, the general solution to the set
of equations 
${\cal H}_{g x}={\cal E}_{g xx}={\cal E}_{g yy}={\cal E}_{g zz}=0$ is 
%
\begin{equation}
 \beta = \beta_0(t),
\end{equation}
where $\beta_0(t)$ is an arbitrary function of $t$~\footnote{
With $\lambda=1$, we would obtain
$\beta=\beta_{IR}\equiv\beta_0(t)+x/(t_c-t)$,  where $\beta_0(t)$ is an
arbitrary function of $t$ and $t_c$ is a constant. By using the residual
symmetry (\ref{eqn:residual-symmetry}) we can set
$\beta_{IR}=x/(t_c-t)$. This would diverge as $t\to t_c$ and thus 
represents a would-be caustics. However, as already stated, near a 
would-be caustics, the system enters the UV regime and $\lambda$ should
deviate from $1$ by RG flow (see footnote
\ref{footnote:deviationoflambda}). Thus, this solution is invalidated
before actually reaching $t=t_c$.}. Thus, by using the residual symmetry 
(\ref{eqn:residual-symmetry}) we can set 
%
\begin{equation}
 \beta = 0. 
\end{equation}
This shows that there is no caustics with codimension one.

\section{Dark matter as integration constant}
\label{sec:darkmatter}

As already mentioned, in Ho\v{r}ava-Lifshitz gravity there is
no local Hamiltonian constraint. This might cause worries since in
general relativity the local Hamiltonian constraint is nothing but the
Poisson equation, which is one of the most important equations for
gravity. Actually, in the following we shall see that the absence of
local Hamiltonian constraint leads to an interesting consequence. A
component which behaves like dark matter emerges after solving the 
system of equations when we interpret general solutions. We will see
that the Poisson equation with ``dark matter'' built-in is satisfied by
the solutions.

Now, following ref.~\cite{Mukohyama:2009mz}, let us define energy
density $\rho_d$ of ``dark matter as integration constant'' by 
%
\begin{equation}
 \rho_d \equiv  -\frac{M_{Pl}^2}{2}(K^{ij}p_{ij}-R-L_{z>1})
  - T_{\mu\nu}n^{\mu}n^{\nu}. 
\end{equation}
Consistency with (\ref{eqn:conservation}) requires that 
%
\begin{equation}
 \partial_t \int dx^3 \sqrt{g}\rho_d = 0, 
\end{equation}
but this is automatically satisfied because of the Hamiltonian
constraint. Equations of motion in Ho\v{r}ava-Lifshitz gravity are
summarized as 
%
\begin{eqnarray}
 M_{Pl}^2\tilde{G}^{(4)\mu\nu} & = & 
  T^{\mu\nu} + \rho_d n^{\mu}n^{\nu}, 
  \label{eqn:modifiedEinstein}\\
 \int dx^3 \sqrt{g}\rho_d & = & 0, 
  \label{eqn:HamiltonianConstraintRhod}
\end{eqnarray} 
where
%
\begin{equation}
 M_{Pl}^2\tilde{G}^{(4)\mu\nu}
  = -{\cal H}_{g\perp}n^{\mu}n^{\nu}
  + {\cal H}_{g i}g^{ij}
  \left[n^{\mu}\left(\frac{\partial}{\partial x^j}\right)^{\nu}
   + n^{\nu}\left(\frac{\partial}{\partial x^j}\right)^{\mu}
  \right]
  - {\cal E}_{g ij}g^{ik}g^{jl}
  \left(\frac{\partial}{\partial x^k}\right)^{\mu}
  \left(\frac{\partial}{\partial x^l}\right)^{\nu}.
\end{equation}
By taking divergence of (\ref{eqn:modifiedEinstein}), we obtain
%
\begin{equation}
 (\partial_{\perp}\rho_d + K\rho_d)n_{\mu} 
  = -\nabla^{\nu}( T_{\mu\nu}-M_{Pl}^2\tilde{G}^{(4)}_{\mu\nu}),
\end{equation}
where $\partial_{\perp}=n^{\mu}\partial_{\mu}$. Because of the spatial
diffeomorphism invariance, the right hand side is proportional to
$n_{\mu}$. Thus, this equation has only one  non-vanishing component. By
contracting with $n^{\mu}$, we obtain the (non-)conservation equation of
``dark matter''~\cite{Mukohyama:2009mz}:
%
\begin{equation}
 \partial_{\perp}\rho_d + K\rho_d = n^{\mu}\nabla^{\nu}
  [T_{\mu\nu}-M_{Pl}^2\tilde{G}^{(4)}_{\mu\nu}].
  \label{eqn:non-conservation}
\end{equation}
In the IR limit with $\lambda\to 1$, $\tilde{G}^{(4)\mu\nu}$ reduces to
the $4$-dimensional Einstein tensor $G^{(4)\mu\nu}$ and
eq.~(\ref{eqn:modifiedEinstein}) reduces to the Einstein equation with
``dark matter'' 
%
\begin{equation}
 M_{Pl}^2G^{(4)\mu\nu} = 
  T^{\mu\nu} + \rho_d n^{\mu}n^{\nu}. 
\end{equation}
From this equation, one can obtain the Poisson equation with ``dark
matter'' built-in. In the same limit, (\ref{eqn:non-conservation})
reduces to the conservation equation,
%
\begin{equation}
 \partial_{\perp}\rho_d + K\rho_d = 0.
\end{equation}

Note that $\rho_d$ can be positive everywhere in our patch of the
universe. In a homogeneous spacetime such as the FRW spacetime, the
global Hamiltonian constraint is as good as local one since all spatial
points are equivalent. However, in inhomogeneous spacetimes there are 
drastic differences. If the whole universe is much larger than the
present Hubble volume then it is natural to expect that the universe far
beyond the present Hubble horizon is different from our patch of the
universe inside the horizon. In this case, the 
global Hamiltonian constraint (\ref{eqn:HamiltonianConstraintRhod}) does
not restrict the universe inside the horizon. Even if we approximate our
patch of the universe inside the present horizon by a FRW spacetime,
the whole universe can include inhomogeneities of super-horizon scales
and, thus, the global Hamiltonian constraint does not restrict the FRW
spacetime which just approximates the behavior inside the
horizon. Therefore, $\rho_d$ can be positive everywhere in our patch of
the universe inside the present Hubble horizon.

In the UV epoch, fluctuations of matter fields as well as metric
fluctuations act as the source term in
(\ref{eqn:non-conservation}). Since such fluctuations include modes with
various wavelengths, fluctuations of ``dark matter'' are generated with
various wavelengths, including those far longer than the size
corresponding to the present Hubble horizon. Therefore, there must
certainly be large enough regions with positive $\rho_d$. In principle
it should be possible to predict a typical amplitude of $\rho_d$, once a
model of the early universe is specified in the context of
Ho\v{r}ava-Lifshitz gravity.

Note also that we do not have to promote the ``dark matter as
integration constant'' to an independent dynamical field as far as the
scalar graviton, the tensor graviton and matter fields are considered as
independent dynamical fields in the initial value formulation. The
initial value formulation of Ho\v{r}ava-Lifshitz gravity consists of
dynamical equation (\ref{eqn:DynamicalEquation}), global Hamiltonian
constraint (\ref{eqn:HamiltonianConstraint}), local momentum constraint
(\ref{eqn:MomentumConstraint}) and gauge conditions. Of course, the
constraints are preserved by the dynamical equation. In this language,
the ``dark matter'' emerges only after solving the system of equations
when we try to interpret a solution.

In order to describe the scalar graviton, the commonly used method
called St\"{u}ckelberg formalism does not seem to be useful here. If we
adopted it to construct an effective field theory of the scalar
graviton then, unlike ghost condensate~\cite{ArkaniHamed:2003uy}, the
foliation-preserving diffeomorphism would forbid $h_{00}^2$ and thus
$\dot{\pi}^2$. For this reason, this description would not include a
healthy kinetic term even with $\lambda\ne 1$ and would get strongly
coupled at all scales in Minkowski background~\cite{Blas:2009yd}. In
non-vanishing ``dark matter'' backgrounds, the strong-coupling scale of 
the St\"{u}ckelberg field becomes finite but is still as low as
$\rho_d^{1/4}$ even with $\lambda\ne 1$~\cite{Blas:2009yd}. This
indicates breakdown of this description but does not imply inconsistency
of the underlining UV theory. The physical reason for this is that the
St\"{u}ckelberg field has vanishing overlap with the scalar
graviton. Indeed, if we adopt for example the gauge used in
\cite{Horava:2009uw} (without introducing a St\"{u}ckelberg field) then
the scalar graviton has a finite kinetic term away from $\lambda=1$ and
does not exhibit the strong coupling mentioned above for the
St\"{u}ckelberg field. (If we introduce the St\"{u}ckelberg field then
we should use a gauge-invariant variable representing the scalar
graviton and solve constraint equations. After all, we should be able to
obtain a finite kinetic term away from $\lambda=1$ in Minkowski
background.) Of course, even in this description, we have to take into
account nonlinear interactions carefully when we take the limit
$\lambda\to 1$ to see if there is an analogue of the Vainshtein
effect~\cite{Vainshtein:1972sx}. We hope to come back to the issue of
Vainshtein-like effect in the near future.

If we consider a group of many microscopic lumps of ``dark matter'' then
collisions and bounces among those lumps may accumulate to generate
non-trivial effects in macroscopic scales. This is exactly the spirit of
renormalization group. As far as gravity at astrophysical scales is
concerned, this kind of collective behavior of small lumps of ``dark
matter'' might mimic behavior of a cluster of particles with velocity
dispersion. If this is the case then microscopic lumps of ``dark
matter'' play the role of particles in usual dark matter
models. It is certainly interesting to see what happens if a group of 
astrophysically large number of such microscopic lumps of ``dark
matter'' collides with another group. Clearly, detailed investigation is
necessary to understand rich dynamics of ``dark matter'' from
microscopic to macroscopic scales.

The foliation-preserving diffeomorphism invariance is the fundamental
symmetry of the theory. Thus, the whole system including all matter
fields must respect this symmetry. In particular, the matter action must
be invariant under the $3$-dimensional spatial diffeomorphism. This
fact, combined with the orthogonality of the flow of ``dark matter'' to
the constant time hypersurface, means that the dispersion relation for
each matter field, 
%
\begin{equation}
 \omega^2 = \frac{1}{M^{2z-2}}k^{2z}
  + \cdots + c_s^2k^2 + m^2, 
\end{equation}
is defined in the rest frame of ``dark matter''. Thus, in the region
where matter fields move relative to ``dark matter'' with large relative
velocities, higher order terms in the dispersion relation of matter
fields can become important. It is certainly interesting to investigate
astrophysical implications of such effects.

\section*{Note added}

The issue of caustics in Ho\v{r}ava-Lifshitz gravity with the
projectability condition was discussed in \cite{Blas:2009yd}. Following
recommendation by an anonymous referee, here we would like to comment on
it. Ref.~\cite{Blas:2009yd} indeed has three statements about
Ho\v{r}ava-Lifshitz gravity with the projectability condition: (i) "dark
matter" forms caustics; (ii) "dark matter" is described by ghost
condensate~\cite{ArkaniHamed:2003uy}; (iii) the scalar sector gets
strongly coupled at the scale $\Lambda\sim\rho_d^{1/4}$ even with
$\lambda\ne 1$. Actually, these three comments are not correct for the 
following reasons. (i) They did not take into account repulsive gravity
due to nonlinear higher curvature terms as mentioned in
Sec.~\ref{sec:no-caustic} of the present paper. (ii) Ghost condensate 
and Ho\v{r}ava-Lifshitz gravity have different symmetries as mentioned
in the third-to-the-last paragraph of Sec.~\ref{sec:darkmatter} of the
present paper. (iii) The strong coupling away from $\lambda=1$
found in \cite{Blas:2009yd} indicates sickness of their description,
i.e. the way dynamical degrees of freedom are identified, but does not 
imply inconsistency of the underlining UV theory. For this point, see
the fourth-to-the-last and the third-to-the-last paragraphs of
Sec.~\ref{sec:darkmatter} of the present paper.

\section*{Acknowledgements}

The author would like to thank Diego Blas, Andrei Frolov, Oriol Pujolas,
Marco Serone, Sergey Sibiryakov, Takahiro Tanaka and Chul-Moon Yoo for
useful discussions. The work of the author was supported in part by MEXT 
through a Grant-in-Aid for Young Scientists (B) No.~17740134, by JSPS
through a Grant-in-Aid for Creative Scientific Research No.~19GS0219,
and by the Mitsubishi Foundation. This work was supported by World
Premier International Research Center Initiative (WPI Initiative), MEXT, 
Japan.


\end{document}